 \documentstyle[prl,aps,multicol]{revtex}

\begin{document} \draft
 \title{Transfer Matrix DMRG for Thermodynamics of
 One-Dimensional Quantum Systems } \author{Xiaoqun Wang$^1$
 and Tao Xiang$^2$} \address{$^1$Institut Romand de
 Recherche Num\'erique en Physique des Materiaux,
 IN-Ecublens, CH-1015 Lausanne, Switzerland}
 \address{$^2$Interdisciplinary Research Centre in
 Superconductivity, The University of Cambridge, Cambridge,
 CH3 0HE, United Kindom} \date{\today} \maketitle
 \begin{abstract} The transfer matrix DMRG method for one
 dimensional quantum lattice systems has been developed by
 considering the symmetry property of the transfer matrix
 and introducing the asymmetric reduced density matrix.  We
 have evaluated a number of thermodynamic quantities of the
 anisotropic spin-1/2 Heisenberg model using this method and
 found that the results agree very accurately with the exact
 ones.  The relative errors for the spin susceptibility are
 less than $10^{-3}$ down to $T=0.01J$ with 80 states kept.

\end{abstract} 
\begin{multicols}{2} 

The density matrix renormalization
group (DMRG) \cite{white} is a powerful numerical method for
studying the ground and low-lying state properties of low
dimensional lattice models.  It has been applied
successfully to a number of strongly correlated systems at
zero temperature in one dimension (1d)
\cite{one,Sorensen,Bursill,MK,Hallberg} as well as two
dimensions (2d) \cite{Liang,Xiang,White96}.  In 1995,
Nishino pointed out that as the partition function of a 2d
classical system is determined only by the maximum
eigenvalue of a transfer matrix in the thermodynamic limit,
the DMRG idea can be extended to find this extreme
eigenvalue and the corresponding eigenstate.  Subsequently
one can study thermodynamic properties.  He calculated the
specific heat of the 2d Ising model using this so called
transfer matrix DMRG method and found that the result agrees
very accurately with the exact solution\cite{Nis}.
Recently, Bursill, Xiang and Gehring have developed a
transfer matrix DMRG algorithm for 1d quantum
systems\cite{Rob}.  They tested the method on the dimerized
spin $1/2$ XY model and obtained encouraging results.
However, a proper implementation of the transfer matrix DMRG
for studying the thermodynamic properties of 1d quantum
systems, especially at low temperature, remains challenging.

In this paper, we report our recent progress on the
development of the transfer matrix DMRG method for 1d
quantum systems.  We have considered the symmetry properties
of the quantum transfer matrix\cite{Suzuk,Betsu} and
introduced an asymmetric reduced density matrix which
optimizes truncated basis states.  This study allows us to
achieve significantly accurate results at low temperature
and greatly enhances the applicability of the method.  For
the $S=1/2$ Heisenberg antiferromagnetic chain, we found
that the relative error for the spin susceptibility is of
the order of $10^{-4}$ down to the temperature $T=0.01J$
($J$ is the exchange constant) with $80$ states kept.  This
size of error and the value of the temperature are smaller
than those for typical Quantum Monte Carlo results as well
as the thermodynamic DMRG results\cite{Moukouri}.  
The thermodynamic DMRG method\cite{Moukouri} cannot 
treat accurately a physical system where the correlated 
length diverges (for example the low temperature 
region of the S=1/2 Heisenberg model) because of the 
finte size effect. However, there is no such problem in 
the transfer matrix DMRG method since this method 
deals directly with an infinite lattice system.

We demonstrate our method using the 1d anisotropic
spin-$1/2$ Heisenberg antiferromagnetic model:
\begin{equation} \hat H= \sum_{i}^{N}\hat h_i;~\hat
h_i=J\left[\hat S_i^x\hat S_{i+1}^x+\hat S_i^y\hat S_{i+1}^y
+\Delta\hat S_i^z\hat S_{i+1}^z \right].  \end{equation} We
shall set $J=1$.  To apply the DMRG idea, we use the Trotter
formula to decompose the partition function.  $\hat H$ is
separated into two parts, $\hat H_o$ and $\hat H_e$,
containing those terms with $i$ being {\it o}dd or {\it
e}ven, respectively.  The partition function is represented
in terms of the quantum transfer matrix ${\cal
T}_M$\cite{Suzuk,Betsu}:  \begin{equation} Z
=\lim_{\epsilon\rightarrow0}{\rm Tr}\left[ {\rm
e}^{-\epsilon \hat H_o} {\rm e}^{-\epsilon \hat
H_e}\right]^M =\lim_{\epsilon\rightarrow0}{\rm Tr} {\cal
T}_M^{N/2} \end{equation} where $\epsilon=\beta/M$,
$\beta=1/T$ and $M$ is the Trotter number.  The elements of
the asymmetric matrix ${\cal T}_M$ are determined by the
product of $2M$ local transfer matrices \begin{eqnarray}
&&\langle \sigma^3_1\cdots\sigma^3_{2M}| {\cal T}_M |
\sigma_1^1\cdots\sigma^1_{2M} \rangle \label{transf}\\ &=&
\sum_{\{\sigma^2_k\}} \prod_{k=1}^M \tau (\sigma^3_{2k-1}
\sigma^3_{2k} |\sigma^2_{2k-1}\sigma^2_{2k})
\tau(\sigma^2_{2k}\sigma^2_{2k+1}
|\sigma^1_{2k}\sigma^1_{2k+1}) \nonumber \end{eqnarray} with
$\tau(\sigma^2_{2M}\sigma^2_{2M+1}
|\sigma^1_{2M}\sigma^1_{2M+1})=\tau(\sigma^2_{2M}\sigma^2_{1}
|\sigma^1_{2M}\sigma^1_{1})$, imposing periodic boundary
conditions in the Trotter direction.  The local transfer
matrix is given by $\tau(\sigma^{i+1}_k \sigma^{i+1}_{k+1}
|\sigma^i_k \sigma^i_{k+1})=$$ \langle s^{i+1}_{k+1},
s^i_{k+1}| \exp(-\epsilon \hat h_i) |s^i_k,s^{i+1}_k\rangle
$ where $\sigma^i_k$$=$$(-1)^{i+k}s^i_k$ and $|s^i_k\rangle$
is an eigenstate of $\hat S_i^z$:  $\hat
S_i^z|s^i_k\rangle$$ =$$s^i_k |s^i_k\rangle$.  The basis
$\left|\sigma^i_{k}\right>\otimes\left|
\sigma^i_{k+1}\right>$ is used to represent $\tau$ and to
construct the corresponding Trotter space.  The superscripts
and subscripts in ${\cal T}_M$ and $\tau$ represent the
coordinates of spins in the real and Trotter space,
respectively.

The local Hamiltonian $\hat h_i$ conserves the total spin at
sites $i$ and $i+1$, i.e.  $s^i_k+s^{i+1}_k=s^i_{k+1}
+s^{i+1}_{k+1}$.  In the Trotter space this conservation law
can be expressed as $\sigma^i_k+\sigma^i_{k+1} =
\sigma^{i+1}_k+\sigma^{i+1}_{k+1}$.  This means that
$\tau(\sigma^{i+1}_k\sigma^{i+1}_{k+1} |\sigma^i_k
\sigma^i_{k+1})$ is block diagonal for each value of
$\sigma^i_k + \sigma^i_{k+1}$.  It turns out that the total
sum of $\sigma^i_k$ at site $i$ is conserved in ${\cal
T}_M$, i.e.  $\sum_k\sigma^i_k=\sum_k\sigma^{i+1}_k$, and
${\cal T}_M$ is block diagonal according to the value of
$\sum_k\sigma^i_k$.  For Eq.  (1), it can be further proved
that the maximum eigenvalue of ${\cal T}_M$ occurs in the
$\sum_k \sigma^i_k=0$ subblock\cite{Koma}.  Therefore only
the $\sum_k \sigma_k^i=0$ subblock in ${\cal T}_M$ is
considered in our DMRG iterations.  This consideration
allows to keep more basis states in the truncation of the
basis set and to save computer CPU time.

In the limit $N\rightarrow \infty$, one can study the
thermodynamic properties using the maximum eigenvalue
$\lambda$ and corresponding left $\left<\psi^L\right|$ and
right $\left|\psi^R\right>$ eigenvectors of the transfer
matrix ${\cal T}_M$.  The free energy is determined purely
by $\lambda$, $F=-\frac 1{2\beta}\ln \lambda$.  From the
derivatives of $F$ one can in principle calculate the
internal energy $U$, the magnetization $M_z$, the specific
heat $C_v$, the spin susceptibility $\chi$, and other
quantities.  However, as it is difficult to evaluate
accurately a derivative of a function in numerical
calculations, we find that it is better to evaluate $U$ and
$M_z$ directly from $\left<\psi^L\right|$ and
$\left|\psi^R\right>$, and then calculate $C_v$ and $\chi$
from the derivative of $U$ and $M_z$, respectively.  For
instance, $U$$= $$\langle \hat H \rangle_T /N$$ = $$\langle
\hat h_1 \rangle_T$$ =$$ \langle \psi^L| {\cal T}_{h_1}
|\psi^R\rangle / \lambda$, where $\langle \, \rangle_T$ is
the thermal average with respect to the thermodynamic
density matrix $\rho_{th}$$=$$\exp(-\beta\hat H)/Z$ and
$\langle \psi^L| \psi^R\rangle$$=1$ is assumed hereafter.
The definition of ${\cal T}_{h_1}$ is similar to that of
${\cal T}_M$ subject to the decomposition.  Its matrix
elements can be obtained from the right hand side of Eq.
(\ref{transf}) by replacing $\tau(\sigma^2_1\sigma^2_2
|\sigma^1_1\sigma^1_2)$ with
$\tau_{h_1}(\sigma^2_1\sigma^2_2
|\sigma^1_1\sigma^1_2)=\langle s^1_2s_2^2|\hat h_1\exp
(-\tau \hat h_1) |s_1^1s^2_1\rangle$.  Similarly, one can
find out the relation between the magnetization $M_z$$ =$$
\langle \sum_i \hat S_i^z$$ \rangle_T/N$ and the maximum
eigenvectors of ${\cal T}_M$.

In our calculation, we fix $\epsilon$ and increase the chain
length $2M$.  For each $M$, the temperature $T=1/\epsilon
M$.  As $M$ is small, one can find $\lambda$, $\langle
\psi^L|$ and $|\psi^R\rangle$ exactly.  For large $M$ we
extend the DMRG idea to approximately but accurately find
$\lambda$, $\langle \psi^L|$ and $|\psi^R\rangle$ for a
periodic time-slice chain.  Figure 1 shows two
configurations of the superblocks of ${\cal T}_M$.  The
superblock consists of two blocks, which we call
renormalized blocks, in the dashed frames and two time
slices.  The system contains a renormalized block and one
slice.  The rest is thus its environment.  We use $n_s$ and
$n_e$ to label the basis states of the renormalized blocks
in the system and the environment, respectively.  The states
of two time slices are represented by $\sigma_1$ and
$\sigma_2$.  The elements of the right transfer matrix is
denoted by ${\cal
T}_o(\sigma_1'',n_s',\sigma_2'';\sigma_1,n_s,\sigma_2)$ or
${\cal
T}_e(\sigma_1',n_s',\sigma_2'';\sigma_1'',n_s,\sigma_2)$ if
$M={\rm odd}$, or even.  The left transfer matrix can be
obtained by transposing the right one.  Therefore the
superblocks ${\cal T}_M$ is given by:  \end{multicols}
\widetext \begin{eqnarray} {\cal
T}_M(n_e',\sigma_2',\sigma_1',n_s';
n_e,\sigma_2,\sigma_1,n_s)=\left\{
\begin{array}{ll}\displaystyle{\sum_{\sigma_1'',\sigma_2''}}
{\cal
T}_o(\sigma_1'',n_e,\sigma_2'';\sigma_1',n_e',\sigma_2')
{\cal T}_o(\sigma_1'',n_s',\sigma_2'';\sigma_1,n_s,\sigma_2)
&~{\rm for ~} M=odd,\\
\displaystyle{\sum_{\sigma_1'',\sigma_2''}} {\cal
T}_e(\sigma_2,n_e,\sigma_2'';\sigma_1'',n_e',\sigma_2')
{\cal T}_e(\sigma_1',n_s',\sigma_2'';
\sigma_1'',n_s,\sigma_2) &~{\rm for~}M=even.
\end{array}\right.  \end{eqnarray} To grow the chain, we
have the following recursive relations:  \begin{eqnarray}
\begin{array}{l} {\cal T}_e(\sigma_1',\tilde
n_s',\sigma_2'';\sigma_1'',\tilde n_s,\sigma_2)=
\displaystyle{\sum_{\sigma''}}
\tau(\sigma_1',\sigma'|\sigma_1'',\sigma'') {\cal
T}_o(\sigma'',n_s',\sigma_2'';\sigma,n_s,\sigma_2), \\ {\cal
T}_o(\sigma_1'',\tilde n_s',\sigma_2'';\sigma_1,\tilde
n_s,\sigma_2) =\displaystyle{\sum_{\sigma''}}
\tau(\sigma_1'',\sigma''|\sigma_1,\sigma){\cal
T}_e(\sigma',n_s', \sigma_2'';\sigma'',n_s,\sigma_2),
\end{array} \end{eqnarray} which enlarge the renormalized
blocks by one slice with $\left|\tilde
n_s\right>=\left|\sigma\right>\otimes \left|n_s\right>$ as
the corresponding Trotter space and $\left|\sigma\right>$ as
the states of a spin added.  Initially, for $2M=4$, ${\cal
T}_e(\sigma_1',\sigma',\sigma_2'';\sigma_1'',\sigma,\sigma_2)=
\sum_{\sigma''}\tau(\sigma_1',\sigma'|\sigma_1'',\sigma'')
\tau(\sigma'',\sigma_2''|\sigma,\sigma_2)$.  As the number
of states in $\left|\tilde n_s\right>$ exceeds $m$, ${\cal
T}_e(\sigma_1',\tilde n_s',\sigma_2'' ;\sigma_1'',\tilde
n_s,\sigma_2)$ and ${\cal T}_o(\sigma_1'',\tilde
n_s',\sigma_2'' ;\sigma_1,\tilde n_s,\sigma_2)$ are
renormalized by:  \begin{equation} \begin{array}{l} {\cal
T}_e(\sigma_1',n_s',\sigma_2'';\sigma_1'',n_s,\sigma_2)
=\displaystyle{\sum_{\tilde n_s'\tilde n_s}}
O^{l}(n_s',\tilde n_s'){\cal T}_e(\sigma_1',\tilde
n_s',\sigma_2'' ;\sigma_1'',\tilde n_s,\sigma_2) O^r(\tilde
n_s,n_s), \\ {\cal
T}_o(\sigma_1'',n_s',\sigma_2'';\sigma_1,n_s,\sigma_2)
=\displaystyle{\sum_{\tilde n_s'\tilde n_s}}
O^{l}(n_s',\tilde n_s'){\cal T}_o(\sigma_1'',\tilde
n_s',\sigma_2'' ;\sigma_1,\tilde n_s,\sigma_2) O^r(\tilde
n_s,n_s), \end{array} \end{equation} where the
transformation matrices $O^{l,r}$ are constructed by $m$
truncated basis states from a reduced density matrix
discussed below.  Other operators such as $\hat h_1$ can be
renormalized by Eq.  (6) with $\tau_{h_1}$ instead of $\tau$
in Eq.  (5).

 \begin{multicols}{2} \narrowtext 
 
 We compute the maximum
eigenvalue, $\lambda$, and the corresponding right
eigenvector, $|\psi^R\rangle$, of ${\cal T}_M$ using a
projection method.  Iterating $|\psi_K\rangle={\cal
T}_M|\psi_{K-1}\rangle$, we reach $|\psi^R\rangle
=|\psi_K\rangle$ and ${\cal T}_M |\psi^R\rangle=\lambda
|\psi^R\rangle $ for sufficient large $K$.  $|\psi_0\rangle$
is an arbitrary trial vector which is not orthogonal to
$|\psi^R\rangle$.  In our calculations, we find that the
value of $K$ needed for producing an eigenvalue with a
relative error less than $10^{-16}$ is generally less than
20, but it increases with increasing $M$.  The left
eigenvector $|\psi^L\rangle$ can be calculated similarly.
However, for systems as we study here, the wave function of
$\langle \psi^L|$ can be directly read out from the wave
function of $|\psi^R\rangle$:  $\psi^L(n_s,\sigma_2,
\sigma_1,n_e) = \psi^R(n_e,\sigma_2,\sigma_1, n_s)$ by
constructing the superblocks with a reflection symmetry as
involved in Eq.  (4).

A density matrix for the whole system (i.e.  superblock) can
be defined as $\rho = {\cal T}_M^{N/2} / {\rm Tr} {\cal
T}_M^{N/2}$.  This is a generalization of the thermodynamic
density matrix $\rho_{th}$ in the Trotter space.  We form
the reduced density matrix for the augmented renormalized
block by performing a partial trace on $\rho$ for the states
of the environment \begin{equation} \rho_s = \frac { {\rm
Tr}_{n_e\sigma_2} {\cal T}_M^{N/2}} { {\rm Tr} {\cal
T}_M^{N/2}} .  \end{equation} In the thermodynamic limit,
$\rho_s = {\rm Tr}_{n_e\sigma_2} |\psi^R\rangle \langle
\psi^L|$, thus the matrix element of $\rho_s$ is given by
\begin{equation} \rho_s (\tilde n_s', \tilde n_s )=
\sum_{\tilde n_e} \psi^{L}(\tilde n_e ,\tilde n_s')
\psi^{R}(\tilde n_e,\tilde n_s) \end{equation} with $|\tilde
n\rangle=|\sigma\rangle\otimes|n\rangle$.  $\rho_s$ is an
asymmetric matrix since $\langle \psi^L| \not=
(|\psi^R\rangle )^\dagger$.  The eigenvalue of $\rho_s$
gives the probability of the corresponding eigenstates onto
which the system is projected as the response to its
environment.  ($|\psi^R\rangle \langle \psi^R|$ which is
used to define the density matrix for the augmented system
block in Ref.  \cite{Rob} is not a true projection operator
for the maximum eigenvectors of ${\cal T}_M$.)  The
transformation matrices $O^{l,r}$ in Eq.  (6) are thus built
up by using $m$ left or right eigenvectors of $\rho_s$
corresponding to $m$ most probable eigenvalues.

Systematic errors come from two sources.  One is the
finiteness of $\epsilon$, and the other is the truncation of
basis set in the DMRG iterations.  The first type of error
is generally very small and in principle it can be further
reduced by doing an extrapolation with respect to
$\epsilon^2$\cite.  The error due to the truncation is
difficult to estimate.  A lower bound for this type of error
is given by the truncation error, $p_m=1-\sum_i^m w_i$,
where $w_i$ ($i=1$,$\cdots$, $m$) are the $m$ largest
eigenvalues of $\rho_s$.  We found that $p_m$ is generally
less than $10^{-5}$ when $m=16$ and decreases rapidly with
increasing $m$ for the spin 1/2 system.

Figure 2 shows the results for the specific heat $C_v(T)$
down to $T=0.02$ with $m=80$ and $\epsilon=0.05$ for
$\Delta=0, 1$ cases.  $C_v$ is obtained from the first
derivative of $U$.  For the $XY$ model ($\Delta$=0), we find
that the relative errors are less than $10^{-5}$ down to
$T=0.02$ for $U$ and less than $10^{-3}$ down to $T=0.03$
for $C_v$ compared with the exact results.  For the
isotropic antiferromagnetic Heisenberg model ($\Delta$=1),
the precision of the results is similar.  The maximum value
of $C_v$ is 0.3515 at $T=0.47$.  At low temperature $C_v$
varies linearly with $T$.  The coefficient of the $T$ term
is shown to be $2\theta/(3\sin \theta)$ with
$\theta=\cos^{-1} \Delta$\cite{Tak}.  By fitting our results
with a polynomial up to $7th$ order in $T$ for
$0.03\stackrel{<}{\mbox{\tiny $\sim$}}T\leq 0.1$, we found
that the coefficients of the linear terms are $1.041$ and
$0.665$ for $\Delta=0$ and 1, respectively.  The difference
between our results and the exact ones for the linear
coefficients is less than 1\%.

For comparison, we also calculated U and $C_v$ with a
 symmetric density matrix as defined in Ref.  \cite{Rob}.
 At high temperature, the results obtained with a symmetric
 density matrix agree well with those obtained with an
 asymmetric density matrix.  However, at low temperature we
 found that the results obtained with an asymmetric density
 matrix are more accurate than those obtained with a
 symmetric density matrix (Figure 2).  The relative errors
 for $U$ and $C_v$ obtained with the symmetric density
 matrix are generally larger than $10^{-2}$ at low
 temperature.

Figure 3 shows our results for the spin susceptibility
$\chi(T)$ down to $T=0.01$ with $m=80$ and $\epsilon=0.05$
for $\Delta=0,1$ cases.  $\chi(T)$ is obtained from the
first order derivative of $M_z$, which is equal to
$M_z(T,B)/B$ for sufficient small magnetic field since
$M_z(T,0)=0$.  In our calculations, $B=0.003$ is used.  For
both cases, the relative error is less than $10^{-3}$ down
to $T=0.01$.  We note that the results of spin
susceptibility are generally more accurate than those of the
specific heat.  In the inset, we compare numerical results
for $\Delta=1$ with $m=32$ and $80$ to the exact results in
the low temperature regime.  For $m=32$ the relative error
is of the order of $10^{-3}$ at $T=0.01$.  Our results are
systematically better than those obtained by Moukouri and 
Caron with the thermodynamic DMRG method\cite{Moukouri}.

Our computations were performed on DEC Alpha stations.  It
takes about 14000 seconds on a 175MHz station to generate a
superblock size of 2M=4000 for $m=32$.

In conclusion, the quantum transfer matrix DMRG method with 
asymmetric density matrices is developed.  We have
calculated a number of thermodynamic quantities for the
anisotropic Heisenberg antiferromagnetic spin-1/2 model and
found the results agree very accurately with exact ones.
Our investigation shows that the transfer matrix DMRG is a
very promising method for studying thermodynamic properties
of 1d quantum systems.

We are very grateful to R.J.  Bursill, P.  Fulde, G.A.
Gehring, T.  Nishino, I.  Peschel and X.  Zotos for
stimulating discussions, and S.  Eggert and S.  Moukouri for
the helpful correspondence.  X.W.  is supported by the Swiss
National Fond Grand No.  933.62.186.153.

\narrowtext

\begin{figure} \caption { Configurations of the superblocks:
(a) $M=odd$ and (b) $M=even$.  The left and right transfer
matrices are connected, via the summation over states
$\sigma_1''$ and $\sigma_2''$, to form a periodic time-slice
chain for ${\cal T}_M$.}  \end{figure} \begin{figure}
\caption { The specific heat $C_v(T)$ for both $\Delta =0$
and $\Delta =1$.  The solid curve is the exact results.
Circles and pluses are the transfer matrix DMRG results with
the asymmetric and symmetric density matrix, respectively.
$\epsilon=0.05$ and $m=80$ are used in the transfer matrix
DMRG calculations.  Inset:  a polynomial fit for the low
temperature $C_v$.}  \label{fig2} \end{figure}

\begin{figure} \caption { The spin susceptibility $\chi(T)$
for both $\Delta =0$ and $\Delta=1$.  The solid curves and
circles, respectively, are the exact results and the
transfer matrix DMRG results with $\epsilon=0.05$ and
$m=80$.  Inset compares the transfer matrix DMRG results for
$m=32$ (diamonds) and $80$ (circles) with the Bethe ansatz
results in the low temperature regime for $\Delta=1$.}
\label{fig3} \end{figure}

\end{multicols}

\end{document}